\documentclass[a4paper,11pt]{article}
\usepackage{aaskaiid}
\usepackage{orcidlink}
\usepackage{soul}

\title{SKA-VLBI Probes of High-energy Emission Processes in Relativistic Jets}

\author[1]{M. Kadler\orcidlink{0000-0001-5606-6154}}
\ShortName{Kadler et al.} % shortened name list for header 
\author[2,3]{E.K. Bempong-Manful\orcidlink{0000-0002-1727-1224}}
\author[4]{P. G. Edwards}
\author[1,5,6]{F. Eppel\orcidlink{0000-0001-7112-9942}}
\author[1]{C. M. Fromm}
\author[7]{M. Giroletti}
\author[1,5]{J. Heßdörfer\orcidlink{0009-0009-7841-1065}}
\author[8,9,10]{T. Hovatta}
\author[5]{Y.~Y.~Kovalev\orcidlink{0000-0001-9303-3263}}
\author[1]{K. Mannheim}
\author[11]{R. Ojha}
\author[1,5]{F. Rösch \orcidlink{0009-0000-4620-2458}}
\author[5]{E. Ros}

\affiliation[1]{Julius-Maximilians-Universit{\"a}t W{\"u}rzburg, Fakult{\"a}t für Physik und Astronomie, Institut für Theoretische Physik und Astrophysik, Lehrstuhl für Astronomie, Emil-Fischer-Str. 31, D-97074 W{\"u}rzburg, Germany}
\affiliation[2]{Jodrell Bank Centre for Astrophysics, Dept. of Physics \& Astronomy, University of Manchester, Manchester M13 9PL, UK}
\affiliation[3]{School of Physics, University of Bristol, Tyndall Avenue, Bristol BS8 1TL, UK}
\affiliation[4]{CSIRO Space and Astronomy, PO Box 76, Epping, NSW1710, Australia}
\affiliation[5]{Max Planck Institute for Radio Astronomy, Auf dem Huegel 69, 53121 Bonn, Germany}
\affiliation[6]{Joint Institute for VLBI ERIC, Oude Hoogeveensedijk 4, 7991 PD Dwingeloo, The Netherlands}
\affiliation[7]{INAF Istituto di Radioastronomia, Via Gobetti 101, I-40127 Bologna, Italy}
\affiliation[8]{Finnish Centre for Astronomy with ESO (FINCA), Quantum, Vesilinnantie 5, University of Turku, FI-20014 Turku, Finland}
\affiliation[9]{Aalto University Metsähovi Radio Observatory, Metsähovintie 114, FI-02540 Kylmälä, Finland}
\affiliation[10]{Aalto University Department of Electronics and Nanoengineering, PL 15500, FI-00076 Espoo, Finland}
\affiliation[11]{National Aeronautics and Space Administration Headquarters, 300 E St SW, Washington, DC 20546-0002, USA}

\abstract{Relativistic jets {in the nuclei of} active galaxies are ubiquitous sources of high-energy emission. In particular, blazars represent the most luminous persistent X-ray and gamma-ray sources,  {whose defining characteristics are small jet inclination angles to the line of sight. Blazars}  can be detected in many cases up to TeV energies  {and the} largest class of TeV emitting extragalactic  {AGN} is represented by high-synchrotron peaked (HSP) BL Lac objects, which are generally comparably faint radio sources. 
 
 Moreover, evidence has also been accumulated that high-energy cosmic neutrinos detected by IceCube can be associated with blazars. Indeed, bright neutrino emission has long been predicted to be produced in flat-spectrum radio quasars (FSRQs) due to the presence of a strong optical/UV seed photon field allowing interactions with high-energy protons inside the relativistic jet and subsequent pion production and decay. However, the case of TXS0506+056, which is lacking a strong FSRQ-typical broad-line region, has cast  {doubts} on this view. Equally puzzling was the highly significant association of a burst of lower-energy neutrinos with TXS0506+056 during a period of low gamma-ray emission. There is an increasing number of suggested blazar-neutrino associations, along with many cases of coincident flaring radio emission, but in a majority of cases, faint blazars on the level of millijanskies or below have to be considered. 
 
 These high-energy photon and neutrino emission processes hold many unanswered questions including the unknown source of seed-photons for photo-pion production and the infamous Doppler crisis of TeV-emitting BL Lac objects. SKA-VLBI offers the opportunity to achieve superior sensitivity at milliarcsecond resolutions, provided by the combination of the phased SKA-Mid and global VLBI arrays. This opens the possibility to perform high-sensitivity and high-angular resolution imaging and polarimetric probes of faint blazars. 
The resulting high-fidelity spatially resolved parameterizations of structured jets in bright sources will yield key insights to constrain physical models of high-energy photon and particle emission in AGN jets.}

\begin{document}
\maketitle

\section{Introduction}
High-angular resolution observations with next-generation radio facilities will play a key role in high-energy astrophysics by providing sub-milliarcsecond spatially-resolved views of extreme environments in astrophysical sources of gamma-ray and (ultra-) high-energetic particle emissions. 
We are currently seeing the dawn of a new era of this field, which is ushered in by the upcoming launch and upgrades of superior new gamma-ray and neutrino telescopes along with next-generation radio facilities, specifically the Square Kilometre Array (SKA). Here, we discuss the role of the SKA for Very Long Baseline Interferometry (VLBI) observations with increased sensitivity provided by the large collecting area of the phased SKA-Mid array for the study of high-energy emission processes in the relativistic jets of active galactic nuclei (AGNs). For a more general view of AGN jet science within the SKA context, see \cite{Baczko01.2026.SKA}.

The Cherenkov Telescope Array Observatory \citep[CTAO;][]{CTAbook} will offer unprecedented sensitivity and resolution in the TeV gamma-ray regime, and current neutrino detectors like IceCube \citep{IceCube} and KM3Net \citep{KM3NeT} are already detecting a significant number of very-high-energy (VHE) neutrino events of mostly unknown origin. These advances raise crucial questions about the production mechanism of TeV-gamma-rays and VHE neutrinos, which are key to understanding the most energetic processes in the Universe. AGNs are the most powerful and luminous objects in the Universe and have already been shown to emit powerful TeV-gamma-rays and neutrinos. Current major Imaging Air Cherenkov Telescopes have detected VHE emission from $\sim 90$ AGN\footnote{\url{https://tevcat2.tevcat.org}}, with the majority of them being blazars.  {Blazars are} a class of AGN that emit violently variable broadband emission from radio to gamma-ray energies. 
Their canonical spectral-energy-distribution (SED) shows two main emission components:
(1) a low-energy \textsl{hump} generally attributed to synchrotron emission of high-energy electrons, and (2) a {second} \textsl{hump} at higher energies explained by either inverse-Compton scattering of electrons on low-energy photons \citep{Sikora1994} or hadronic interactions due to relativistic protons \citep{Mannheim1993}.
With decreasing luminosity, both peaks are shifted upwards and the high-energy emission can reach the very-high-energy (VHE) regime at TeV-gamma-rays. 
High-peaked BL\,Lac objects (HBLs) are canonically defined as sources whose synchrotron emission hump peaks above $10^{15}$\,Hz \citep{Padovani1995}. In the most extreme cases, they can reach up even higher by up to two orders of magnitude \citep{Ghisellini1999,Biteau2020}. 
{Blazars are of utmost interest for astroparticle physics as possibly dominant sources of ultrahigh-energy cosmic rays and neutrinos} \citep[e.g.,][]{Hillas1984,Mannheim1995}. In particular, HBLs and extreme blazars have been considered in several  studies as relevant neutrino sources \citep[e.g.,][]{Tavecchio2014,Tavecchio2015,Cerutti2015,Padovani2015,Giommi2020}.

Because of their high peak frequencies, HBL blazars are generally faint radio sources and difficult to observe.  
At the same time, radio  data are crucially important to put very-high-energy flaring results into context, as has been demonstrated in previous work \citep[e.g.,][]{Kadler2012,Aleksic2014a,Benke2024a}. 
Radio Doppler factors are surprisingly often found to differ drastically from the Doppler factors derived from high-energy observations \citep{Lister2019,Piner2018}. To explain this so-called \textsl{Doppler crisis of TeV blazars}, models have been proposed involving multiple zones on parsec scales \citep[e.g.,][]{Hervet2019}, which can be investigated with coordinated deep multiwavelength and long-term monitoring observations. Radio variability studies at mid and high frequencies (in the cm- and mm-bands) are of particular importance as they are probing jet activity in  compact jet regions that are expected to have the tightest physical connection to the very-high-energy activity.

\begin{figure}[b!]
\centering
    \includegraphics[width=\textwidth]{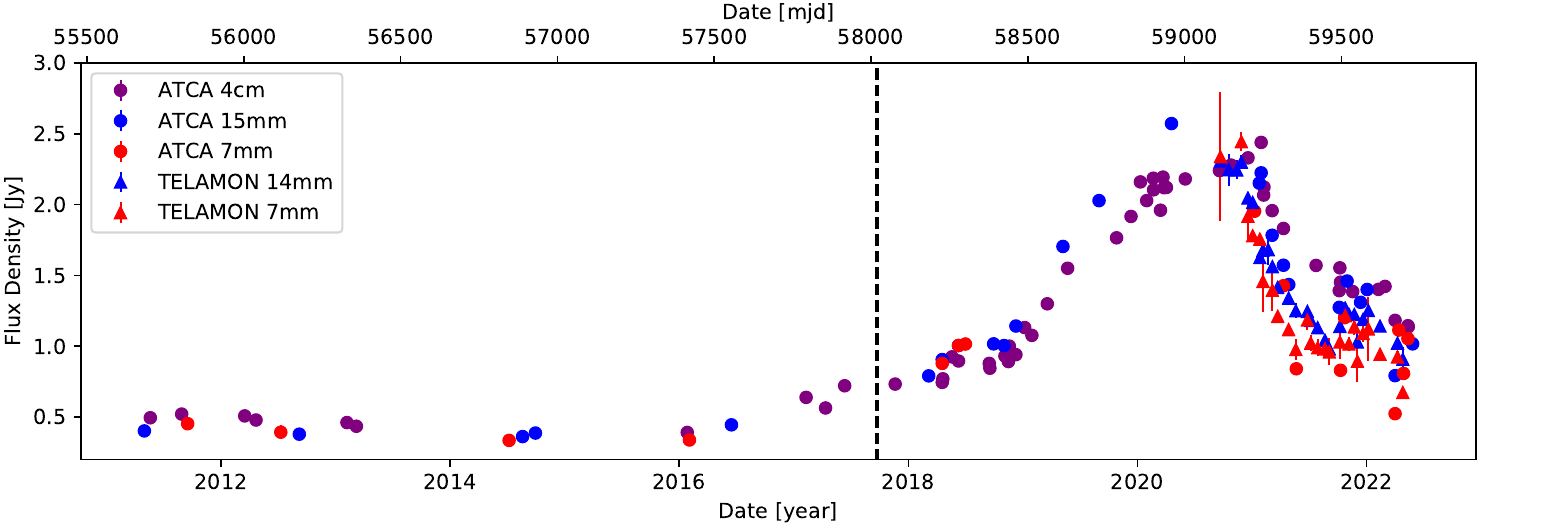}
    \caption{\small \sl Long-term ATCA/Effelsberg radio light curve of TXS\,0506+056 showing the dramatic radio outburst that evolved after the detection of the coincident IC170922A neutrino (dashed line).
    }
    \label{fig:txs0506_lc}
\end{figure}

Recently, it was shown that {radio monitoring programs also play a key role in understanding very-high-energy neutrino emissions}. A tentative picture is emerging in which the detection of very-high-energy neutrino emission might be characteristically associated with radio flares in compact AGNs \citep[see sample studies with significance estimates in][]{Plavin2020,Hovatta2021,2024A&A...690A.111K}. This general behavior has already been seen before in case of the three individual high-confidence neutrino-associated blazars PKS\,1424$-$418 \citep{Kadler2016}, TXS\,0506+056 \citep{Kun2019}, and PKS\,1502+106 \citep{ATel12996}. 
An impressive visualization is shown in Fig.~\ref{fig:txs0506_lc} (cf. \citealt{2024MNRAS.527.8784A}), demonstrating the dramatic radio outburst of TXS\,0506+056 that started around the arrival time of the IC170922A neutrino.

The general correlation between radio-bright AGNs and IceCube neutrinos, initially suggested by \cite{Plavin2020}, remains a debated and actively investigated topic.
Recent sample-based and individual AGN studies indicate that neutrino-selected blazars exhibit higher Doppler-boosting factors \citep{2025ApJ...991...33P,2025A&A...700L..12K,2025arXiv251016585K}.
While some works support a possible physical connection (see the references above), others have challenged these results \citep[e.g.,][]{IceCube2023,Zhou2021}.
It has been shown that only a limited fraction ($\lesssim$10–30\%, depending on model assumptions) of IceCube neutrinos can be associated with the brightest blazars in radio and gamma-ray catalogs \citep{Aartsen2017}.

Evidently, dedicated sensitive radio programs hold an immense detection potential in these new fields connecting radio astronomy with astroparticle physics. In particular, sensitive programs with broad mid-to-high frequency coverage and polarimetric capabilities provide a gold standard as flares often peak earlier and brighter at higher radio frequencies \citep[e.g.,][]{Myserlis2016,Angelakis2019},
and their spectral evolution provides elusive
clues to emission processes over single-frequency data.

\section{Opportunities opened by SKA-VLBI}
VLBI can serve as a magnifying glass for gamma-ray astronomy: it provides  {direct probes of sub-milliarcsecond-scale jet sub-structures that can be associated with high-energy variability and flaring activity in blazars}
\citep[e.g.,][]{Dotson2015,Roesch2025}.
The continuous \textit{Fermi}/LAT GeV gamma-ray light curves and spectra have very successfully been combined with VLBI and broadband monitoring of AGN jets, revealing the statistics of radio--GeV correlations in blazars \citep[e.g.,][]{Jorstad2001b,Krauss2016}
and deep insights into many individual jet systems \citep[e.g.,][]{Abdo2009_NGC1275,Jorstad2013,MacDonald2017}.
Going forward, a similar impact of VLBI is expected on TeV-blazar studies: broadband, quasi-simultaneous data of a large sample of sources in different flaring states are needed to address the salient open questions in AGN jet research. For TeV blazars, fainter source populations have to be targeted
 {in order to increase sample statistics for all relevant sub-samples.}
 {This is what} makes the phased SKA-Mid  {a key} element of future VLBI arrays in this context, especially in the southern hemisphere.

In the pre-SKA era,  most advanced VLBI studies of high-energy emitting AGN in the southern sky have been made within the scope of the TANAMI program \cite{Ojha2010,Mueller2011,Mueller2014a,Mueller2014b,Schulz2016,Mueller2018,Benke2024a} with the Australian Long Baseline Array and associated telescopes. Moreover, the Very Long Baseline Array (VLBA) has been used to investigate TeV blazars with a special focus on the Doppler crisis \citep[see][and references therein]{Piner2018}. These studies have been severely limited by image fidelity due to  baseline and sensitivity constraints. First test observations in 2024 successfully tied the new SKAMPI antenna into the TANAMI array \citep{Wongphecauxson01.2026.SKA}, which demonstrates the high prospects for combined LBA and SKA observations in the future.

\subsection{Characterization of the short-wavelength radio variability patterns of TeV blazars}
\cite{Lindfors2016} have pioneered studies to characterize the radio variability of TeV-emitting blazars based on OVRO 15\,GHz data. They find that simple single-zone emission models cannot explain the variability patterns.  {This underscores the importance of continuous high-sensitivity and densely sampled multi-frequency radio light curves that can separate different jet-emission zones.} With new optimized monitoring programs using more sensitive instruments such as the TELAMON program \citep{Eppel2024} on the Effelsberg 100m telescope, such studies are currently extended to larger and more complete samples. In the southern hemisphere, 15\,GHz  monitoring observations with a phased sub-array of the SKA core could yield even more sensitive data. Dynamic broadband SEDs and the monitoring of polarized emission can yield additional information on emission models including constraints on magnetic-field configurations, particle density and plasma composition \cite[e.g.,][]{Myserlis2016}. 

\subsection{Investigation of jet kinematics and the Doppler crisis in TeV-detected AGNs}
Advanced models proposed to explain the complex observational patterns associated with high-energy emission characteristics of AGN jets often involve multiple zones on parsec scales \cite[e.g.,][]{Hervet2019}. 
Combined spectro-polarimetric flux-density and VLBI monitoring observations are ideally suited to probe such models and investigate the most compelling problems such as the Doppler crisis. 
 {SKA-VLBI has the sensitivity and angular resolution to reveal the putative existence of isolated features within the underlying flow, such as fast (high Doppler-factor) plasma components within otherwise slowly evolving parsec-scale jets}, especially also in linear polarization and related to multi-messenger flares \citep[e.g.,][]{2026A&A...709A..50K}. Continuous monitoring of a large number of Doppler-crisis blazars is needed to detect, parameterize and follow up such events.

The shortest time scales observed (possibly during rare flaring events and often more significant in polarized emission rather than total intensity) are indicative of the highest-relativistic plasma in AGN jets, which makes Band~5b (8.3--15.4~GHz) most important in the SKA context. For SKA-VLBI observations in this band, external  VLBI arrays need to be equipped with matching VLBI backends and compatible technology  \citep{Kadler02.2026.SKA,Bempong-Manful01.2026.SKA}.

\subsection{Locating the emission sites of TeV-gamma rays}
Previous coordinated radio/gamma studies of flat-spectrum radio quasars (FSRQs) show a tight correlation between the activity of the mm-core and high-energy events \cite[see, e.g.,][]{Leon2011,Roesch2025}. This suggests that we often resolve the jet at the distance where the gamma-ray emission is produced. Gamma-ray observations are sensitive to the radiation processes of the highest energy particles in the jet, but do not provide sufficient angular resolution.
The radio morphology of jets can be measured using SKA-VLBI observations for fainter source populations as in the past. This opens up the possibility to study samples of HBLs and extreme blazars to be compared to the previously studied (much brighter) FSRQ source class.

\subsection{SKA-VLBI probes of characteristic jet signatures in total intensity and polarization}

\begin{figure}
    \centering
    \includegraphics[width=\linewidth]{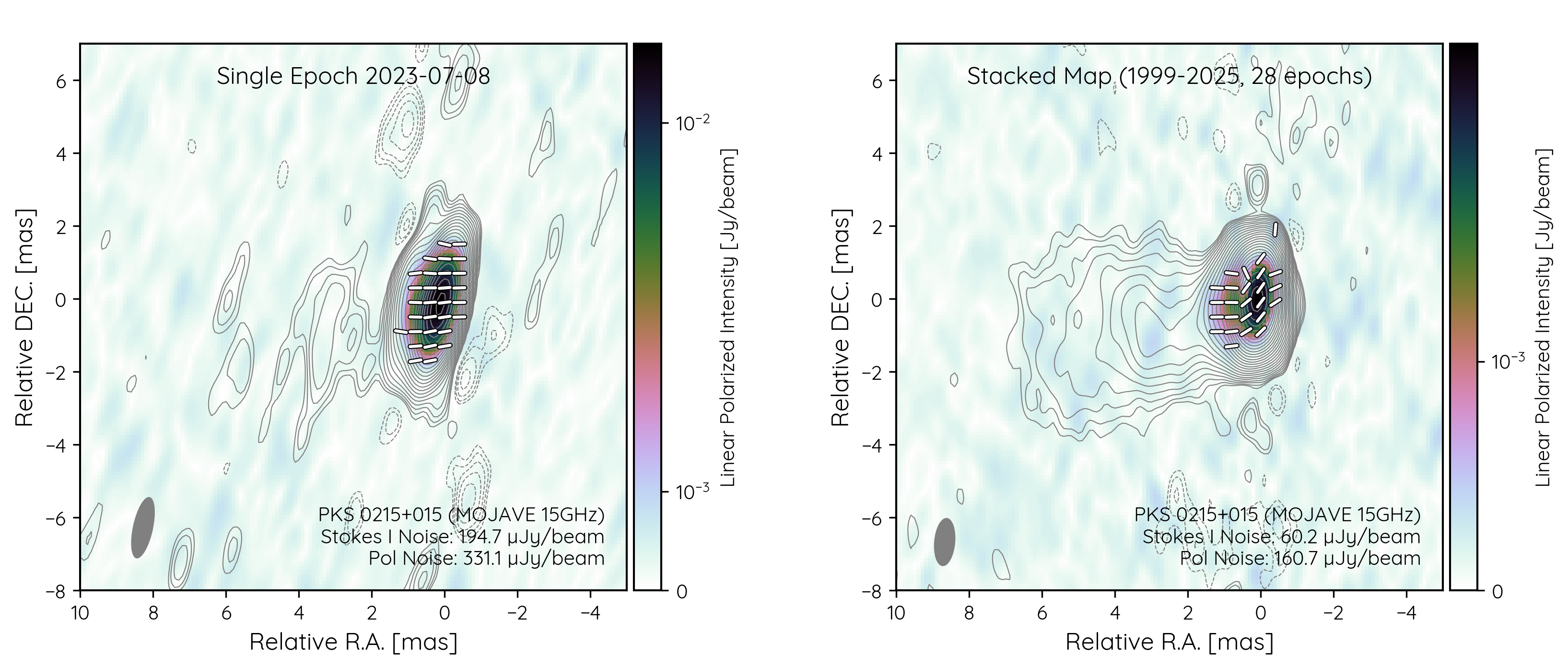}
    \includegraphics[width=\linewidth]{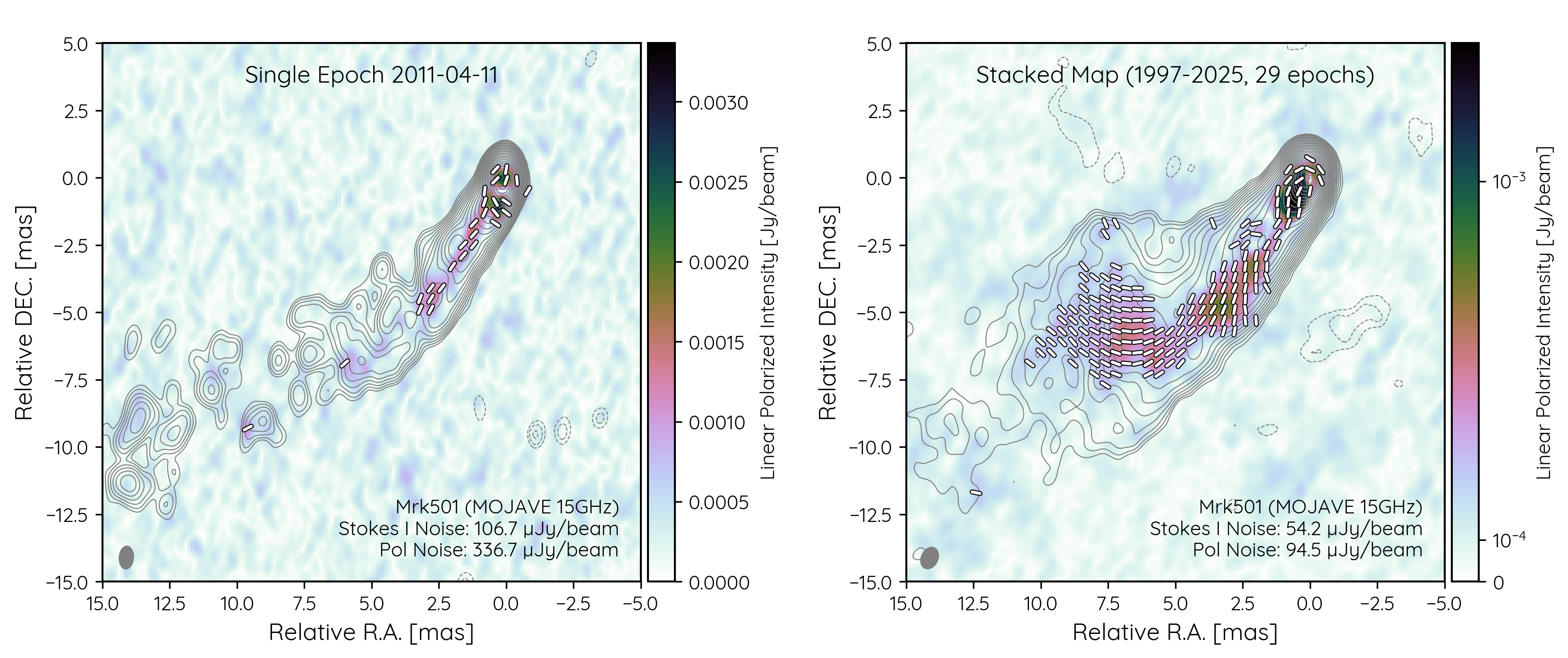}
    \caption{Current instruments have limited sensitivity and can reveal tentative evidence for spine-sheath structures in AGN jets only for the brightest sources. Here, we show selected single-epoch (left column) and stacked maps (right column) of the bright (0.5\,Jy -- 3\,Jy) neutrino-candidate blazar PKS\,0215+015 (top), and the bright (0.6\,Jy -- 1.2\,Jy) extreme blazar Mrk\,501 (bottom), compiled from archival 15\,GHz MOJAVE observations. The contours indicate total intensity, with the lowest contour corresponding to three times the Stokes I noise level, and increasing by a factor of $\sqrt{2}$. The color map shows linear polarized intensity, while the EVPA direction is indicated by the overplotted lines for all regions where the polarized signal is at least three times above its noise.}
    \label{fig:stacked_maps}
\end{figure}

Current models that are used to explain the TeV-emission in AGN involve scenarios such as a spine-sheath structure \cite[e.g.,][]{Ghisellini2005} or the presence of standing recollimation shocks \cite[e.g.,][]{Hervet2019}, which can in principle be resolved through high-resolution VLBI-observations. Previous VLBI-observations have already found hints of such limb-brightened structures in some TeV-detected sources 
\citep[see, e.g.,][]{Piner2009,Ros2020,Janssen2021},
however, the image fidelity and number of these observations is still limited, so that the seed photon field for the hadronic processes cannot fully be resolved and characterized. 
SKA-VLBI will image the jet geometry with unprecedented detail. Moreover, polarized emission can yield additional information on emission models including constraints on magnetic-field configurations, particle density and plasma composition \cite[e.g.,][]{Myserlis2016}. The measurement of polarization signatures of blazars at low radio flux densities is difficult due to their low polarization degrees of typically a few percent at most, as well as Faraday-depolarization in single-dish  programs. Sensitive and well calibrated VLBI data can overcome these limitations \cite[e.g.,][]{Kim2023}.
Moreover, by stacking images from multiple epochs
it is possible to increase sensitivity and reveal extended or faint structure in much more detail than possible with a single-epoch observation --- see Fig.\,\ref{fig:stacked_maps} and also \cite{2025A&A...700L..12K}. 
The downside of this approach is that fast structural variability tends to smear out milliarcsecond-scale jet-component contributions. With the superior sensitivity offered by SKA-VLBI, the dynamic range needed can be obtained in single-epoch observations  {complementary to} epoch-stacking techniques.

\section*{Acknowledgments}
We acknowledge support from the Deutsche
Forschungsgemeinschaft (DFG, grants 434448349, 447572188, 465409577 and 443220636 [FOR5195:
Relativistic Jets in Active Galaxies])
and the European Union (ERC MuSES project 101142396).
This research has made use of data from the MOJAVE database that is maintained by the MOJAVE team \citep{Lister2018}.

\bibliographystyle{abbrvnat-maxbibnames4}
\bibliography{chapter} % if your bibtex file is called example.bib

\end{document}